\documentclass{rspublic}
\begin{document}

\newcommand{\half}{\mbox{$\textstyle \frac{1}{2}$}}
\newcommand{\mba}{\boldsymbol{\alpha}}
\newcommand{\mbb}{\boldsymbol{\beta}}
\newcommand{\mbc}{\boldsymbol{\gamma}}
\newcommand{\mbd}{\boldsymbol{\delta}}
\newcommand{\mbp}{\boldsymbol{\rho}}
\newcommand{\mbq}{\boldsymbol{\sigma}}
\newcommand{\mbr}{\boldsymbol{\tau}}

\title[Theory of Quantum Space-Time]{Theory of Quantum Space-Time}

\author[D.C.~Brody \& L.P.~Hughston]{Dorje~C.~Brody$^*$ and
Lane~P.~Hughston$^\dagger$}

\affiliation{${}^*$Blackett Laboratory, Imperial College, London
SW7 2BZ \\ ${}^\dagger$Department of Mathematics, King's College
London, London WC2R 2LS, UK}

\date{\today}
\maketitle
\input{psfig.sty}
\large

\begin{abstract}{Quantum mechanics, quantum entanglement, causal
spaces, \\ density matrices, cosmological models, hyperspinors \\
Working paper: 29 June 2004} A generalised equivalence principle
is put forward according to which \emph{space-time symmetries and
internal quantum symmetries are indistinguishable before symmetry
breaking}. Based on this principle, a higher-dimensional extension
of Minkowski space is proposed and its properties examined. In
this scheme the structure of space-time is intrinsically quantum
mechanical. It is shown that the causal geometry of such a quantum
space-time possesses a rich hierarchical structure. The natural
extension of the Poincar\'e group to quantum space-time is
investigated. In particular, we prove that the symmetry group of a
quantum space-time is generated in general by a system of
irreducible Killing tensors. When the symmetries of a quantum
space-time are spontaneously broken, then the points of the
quantum space-time can be interpreted as \emph{space-time valued
operators}. The generic point of a quantum space-time in the
broken symmetry phase thus becomes a Minkowski space-time valued
operator. Classical space-time emerges as a map from quantum
space-time to Minkowski space. It is shown that the general such
map satisfying appropriate causality-preserving conditions
ensuring linearity and Poincar\'e invariance is necessarily a
density matrix.
\end{abstract}

\section{Introduction}
\label{sec:1}

The purpose of this paper is to present a novel approach to the
unification of space-time physics and quantum theory. We take the
view that classical space-time itself is not to be regarded as a
primary object which is then subjected to some form of
quantisation procedure.  In our framework the central object is a
mathematical structure that we call a quantum space-time.
Intuitively, this structure can be regarded as the space of all
space-time valued quantum operators. That is to say, each point in
the infinite-dimensional quantum space-time corresponds to a
quantum operator with the property that its expectation, in any
quantum state, is a space-time point. The space of all such
operators has a rich structure that appears to contain all the
elements one needs both for a characterisation of the causal
structure of relativistic space-time as well as a representation
of the phenomena of quantum theory.

Many attempts to unify gravitational physics with other
fundamental forces have pursued the idea of extending
four-dimensional space-time to higher dimensions. Beginning with
the introduction of the Kaluza-Klein theory of gauge potentials,
such extensions have typically been carried out by increasing the
spatial dimension of the space-time, while retaining the special
role played by time. In methodologies of this sort, however, it
cannot be said that quantum mechanical characteristics of the
fundamental forces are adequately incorporated into the structure
of the higher dimensional space-time. Nor can it be said that the
space-time itself is being treated in any useful sense as a
quantum entity. In what follows, however, we demonstrate that if
we take the point of view that the universe is itself
intrinsically quantum mechanical in an appropriate sense, then the
most natural extension of space-time into higher dimensions has a
completely different character from that suggested by the
Kaluza-Klein theory and its generalisations.

The basic idea is as follows. The points of Minkowski space are in
natural correspondence with two-by-two matrices of the form
$x^{{\bf AA}'}$ satisfying a Hermitian condition. Lorentz
transformations are then given by multiplying $x^{{\bf AA}'}$ on
the right and on the left by an element of $SL(2,{\mathbb C})$ and
its complex conjugate, and the Minkowskian metric for the interval
between two points $x^{{\bf AA}'}$ and $y^{{\bf AA}'}$ is obtained
by taking the determinant of their difference. Hermitian matrices
are, on the other hand, familiar objects in quantum mechanics in
their role as physical observables. In quantum mechanics, the
dimensionality of the Hilbert space is directly related to the
dimensionality of the corresponding space of Hermitian operators.
Thus, if quantum theory is to be unified with space-time physics,
it seems natural to extend the space of matrices representing
space-time points to higher dimensions, and thus to assume that
the dimensionality of space-time is much larger than four,
perhaps infinite. A higher-dimensional extension of space-time in
this manner is also consistent with the philosophy often put
forward that spinors (or twistors) are in some respects just as
fundamental, or possibly even more fundamental, than the
space-time points themselves.

The framework we introduce here is motivated by the idea that the
symmetries of space-time and the symmetries of quantum theory are,
at the deeper level, indistinguishable. This generalised
`equivalence principle' reflects the notion that the fundamental
symmetries or approximate symmetries we observe in nature should
have a common origin, and that the breakdown of these symmetries
should also have a common cause. In view of the generalised
equivalence principle we shall therefore postulate in this
investigation that \emph{space-time events are themselves
infinite-dimensional Hermitian matrices}. As in ordinary quantum
mechanics, however, it is both legitimate and desirable to
consider finite dimensional realisations of the framework in some
circumstances. These finite dimensional realisations are given by
spaces of $r$-by-$r$ Hermitian matrices. In this way we obtain for
each $r\geq2$ an $r^2$-dimensional space-time ${\mathcal
H}^{r^2}$. The standard four-dimensional Minkowski space
${\mathfrak M}^{4}$ then emerges as the simplest case.

As we explain later in the paper, we regard such finite
dimensional cases not merely as toy models, but rather as special
situations where a finite-dimensional part of the
infinite-dimensional space-time is (or effectively can be regarded
as) disentangled from the rest of the space-time. In this way the
fundamental role played by the Segr\'e embedding in the geometry
of quantum theory is carried over to the relativistic domain.
Indeed, it is an important feature of our approach that many of
the familiar ideas relevant to the geometry of the quantum state
space are directly applicable to space-time itself, and as such
take on a new physical significance, some aspects of which we
explore in what follows.

The structure of the paper is as follows. In \S2 we introduce the
algebraic formalism appropriate for the manipulation of
$r$-component hyperspinors. The concept of hyperspinors as a
natural higher dimensional generalisation of the familiar
two-component spinors of relativity theory and as a basis for
higher-dimensional space-time theories was introduced by
Finkelstein (1986) and Finkelstein {\it et al}. (1987), and we
build on that work here. The $r^2$-dimensional quantum space-time
${\mathcal H}^{r^2}$ then arises as the tensor product of the
space of $r$-dimensional hyperspinors with its complex conjugate.
In \S3 we investigate the causal structure of ${\mathcal
H}^{r^2}$. This structure is shown to arise by virtue of a
resolution of the vector separating any two points in ${\mathcal
H}^{r^2}$ into a canonical form involving a sum of terms, each of
which can be expressed as a product of a hyperspinor with its
complex conjugate, together with a plus or minus sign. In
particular, we can introduce the concept of future and past
pointing time-like and null vectors in ${\mathcal H}^{r^2}$. This
structure exists despite the fact that ${\mathcal H}^{r^2}$ does
not possess a pseudo-Riemannian metric. Instead, ${\mathcal
H}^{r^2}$ possesses a chronometric form of rank $r$, which induces
a pseudo-Finslerian geometry on ${\mathcal H}^{r^2}$. In \S4 we
look at the variational problem for determining the geodesic
between two time-like separated points, and show that this reduces
to an appropriate linear expression, despite the fact that the
chronometric form is itself a polynomial of degree $r$ in the
separation vector for the given points.

In \S5-6 we study the higher dimensional analogue of the
Poincar\'e group that acts on ${\mathcal H}^{r^2}$. We prove that
the symmetries of this space are generated by a system of $3r^2-2$
irreducible Killing tensors, each of rank $r-1$. Since ${\mathcal
H}^{r^2}$ does not have a pseudo-Riemannian structure for $r>2$,
the link between symmetries and Killing vectors is lost in
general, and is replaced by this more subtle manifestation of
symmetry. We show that the conserved quantities associated with
the hyper-Poincar\'e group can be obtained in terms of algebraic
expressions formed from the Killing tensors. We also derive
appropriate hyper-relativistic generalisations of the familiar
expressions for the momentum, angular momentum, mass, and spin of
a relativistic system.

In \S7 we make a detailed study of the algebraic geometry of the
complex light-cone at a point of ${\mathcal H}^{r^2}$, and examine
the structures arising for various values of $r$. We show that the
space of complex light-like directions is a complex hypersurface
${\mathfrak N}^{r^2-2}$ of degree $r$ in the complex projective
space ${\mathbb P}^{r^2-1}$. The hypersurface ${\mathfrak
N}^{r^2-2}$ can be completely characterised by the fact that it
admits a special hyperspinorial subvariety of the form ${\mathbb
P}^{r-1}\times{\mathbb P}^{r-1}$ . In \S8 we present a higher
dimensional analogue of the Klein representation, and show how
${\mathcal H}^{r^2}$, when complexified and compactified, can be
represented as the Grassmannian of complex $(r-1)$-planes in
${\mathbb P}^{2r-1}$. We show that the causal relations between
points of ${\mathcal H}^{r^2}$ can be understood in terms of the
intersection properties of the corresponding $(r-1)$-planes.

In \S9 we show how the conformal symmetry of the geometry of the
generalised Klein representation can be reduced to the
hyper-Poincar\'e group by the introduction of elements determining
the structure of infinity for this space. The range of
possibilities for structure at infinity is considerably larger
than it is for four-dimensional space-time. As a consequence, as
we show in \S10, the choice of structure at infinity can also give
rise to interesting classes of cosmological models. We point out
that there is a mechanism within our framework whereby the same
structures at infinity responsible for the reduction of the
symmetry of space-time can also be responsible for the breaking of
microscopic symmetries.

In \S11 we further explore the notion of symmetry breaking, and
introduce the idea of the Segr\'e embedding as the basis of the
mechanism according to which space-time degrees of freedom can be
disentangled from microscopic or internal degrees of freedom.
According to this scheme the dimension of the hyperspinor space is
assumed to be even, and each hyperspinor index $A$ with the range
$A = 1,2,\ldots,2n$ is regarded as a clump consisting of a
conventional two-component spinor index ${\bf A} = 1,2$ and an
``internal'' index $i = 1,2,\ldots,n$ . It follows as a
consequence of this symmetry breaking scheme that the points of
${\mathcal H}^{4n^2}$ can be interpreted as \emph{space-time
valued operators}. This is the sense in which ${\mathcal
H}^{4n^2}$ can be regarded as a quantum space-time. Finally, in
support of this interpretation, in \S12 we consider maps from
${\mathcal H}^{4n^2}$ to Minkowski space ${\mathcal H}^{4}$. We
show that if such a map $\rho$ is in a suitably defined sense (i)
linear, (ii) Poincar\'e invariant, and (iii) causal, then $\rho$
is a density matrix, and the map can be interpreted as the
expectation. Thus within our scheme an important element of the
probabilistic interpretation of quantum theory arises as an
emergent property deriving from the causal nature of space-time.

\section{Hyperspinors}
\label{sec:2}

What we aim for in this investigation is not a higher dimensional
pseudo-Riemannian analogue of four-dimensional space-time, but
rather a geometry of a different character that, although richer
in structure than Minkowski space, nevertheless retains a definite
relation to that space---and as a consequence is in a position to
embrace the description of physical phenomena, albeit in a new
setting. We find it convenient to proceed in stages. The first
step is to introduce a special type of causal geometry for which
the space is of dimension $N=r^2$, where $r\geq2$ is an integer.
Later we shall specialise to the case for which $r$ is even, and
introduce some additional structure that cements the relationship
of the higher dimensional space to ordinary Minkowski space. We
shall demonstrate that when $r$ is even there exists a natural
symmetry breaking mechanism that embeds Minkowski space in the
higher dimensional space. By virtue of the geometry of this
embedding we can assign physical properties to elements of the
higher-dimensional space.

For general $r$ we refer to this space as ${\mathcal H}^{N}$, the
quantum space-time of dimension $N=r^2$. If we fix a point of
origin in ${\mathcal H}^{N}$, then a point of ${\mathcal H}^{N}$
can be characterised by its position vector with respect to that
origin. In Minkowski space ${\mathfrak M}^{4}$ such vectors are
naturally isomorphic to elements of the vector space obtained by
taking the tensor product of a complex vector space ${\mathbb
S}^{\bf A}$ with its complex conjugate ${\mathbb S}^{{\bf A}'}$.
We recognise ${\mathbb S}^{\bf A}$ and ${\mathbb S}^{{\bf A}'}$ as
the spaces of unprimed and primed two-component spinors,
respectively. For two-component spinors we use bold upright Roman
indices, and we adopt the usual conventions for raising and
lowering indices, and for complex conjugation. Thus if
$\alpha^{\bf A}\in{\mathbb S}^{\bf A}$, then we write $\alpha^{\bf
A}\epsilon_{\bf AB}=\alpha_{\bf B}$ and $\alpha^{\bf
A}=\epsilon^{\bf AB}\alpha_{\bf B}$, where $\epsilon_{\bf AB}=-
\epsilon_{\bf BA}$. Likewise, for the complex conjugation map we
write $\alpha^{\bf A}\to {\bar\alpha}^{{\bf A}'}$ where
${\bf\alpha}^{{\bf A}'}\in{\mathbb S}^{{\bf A}'}$. A special
feature of the two-component spinor algebra is that the epsilon
spinor functions as a nondegenerate symplectic form that can be
used to establish a linear map from the spin space ${\mathbb
S}^{\bf A}$ to its dual space ${\mathbb S}_{\bf A}$. For further
details of the two-component spinor algebra, see, e.g., Pirani
(1965), Penrose (1968), Penrose \& Rindler (1984,1986).

Our model of quantum space-time generalises the two-component
spinor formalism and its relation to Minkowski space by allowing
the underlying spin spaces to be higher dimensional complex vector
spaces, the elements of which, using the terminology of
Finkelstein (1986) we call \emph{hyperspinors}. Let us write
${\mathbb S}^{A}$ and ${\mathbb S}^{A'}$, respectively, for the
complex $r$-dimensional vector spaces of unprimed and primed
hyperspinors. For hyperspinors we use unprimed and primed italic
indices. It is assumed that these two spaces admit an anti-linear
isomorphism under the operation of complex conjugation. Thus if
$\alpha^{A}\in{\mathbb S}^{A}$, then under complex conjugation we
have $\alpha^{A}\to {\bar\alpha}^{A'}$, where ${\bar\alpha}^{A'}
\in {\mathbb S}^{A'}$. In a standard basis this map conjugates
$\alpha^{A}$ component by component to give the components of
${\bar\alpha}^{A'}$.

The dual spaces associated with the hyperspin spaces ${\mathbb
S}^{A}$ and ${\mathbb S}^{A'}$ will be denoted ${\mathbb S}_{A}$
and ${\mathbb S}_{A'}$, respectively. If $\alpha^{A}\in{\mathbb
S}^{A}$ and $\beta_{A}\in{\mathbb S}_{A}$, then their inner
product is $\alpha^{A}\beta_{A}$. Likewise if $\gamma^{A'}\in
{\mathbb S}^{A'}$ and $\delta_{A'}\in {\mathbb S}_{A'}$ then for
their inner product we write $\gamma^{A'}\delta_{A'}$.

We also introduce the totally antisymmetric hyperspinors of rank
$r$ associated with the spaces ${\mathbb S}^{A}$, ${\mathbb
S}_{A}$, ${\mathbb S}^{A'}$, and ${\mathbb S}_{A'}$. These will be
denoted $\varepsilon^{AB\cdots C}$, $\varepsilon_{AB\cdots C}$,
$\varepsilon^{A'B'\cdots C'}$, and $\varepsilon_{A'B'\cdots C'}$.
The choice of these antisymmetric hyperspinors is canonical up to
an overall scale factor. Once a specific choice has been made for
$\varepsilon_{AB\cdots C}$, then the other epsilon hyperspinors
are determined by the relations $\varepsilon^{AB\cdots C}
\varepsilon_{AB\cdots C}=r!$, $\varepsilon^{A'B'\cdots C'}
\varepsilon_{A'B'\cdots C'}=r!$, and $\varepsilon_{A'B'\cdots C'}=
{\bar\varepsilon}_{A'B'\cdots C'}$, where
${\bar\varepsilon}_{A'B'\cdots C'}$ is the complex conjugate of
$\varepsilon_{AB\cdots C}$. If we introduce a standard basis then
it is convenient to set $\varepsilon_{\mathfrak{ 12}\cdots r}=1$,
which is sufficient to fix the remaining components of the epsilon
hyperspinors. The arguments that follow, however, do not depend on
a specific choice of scale.

The epsilon hyperspinors play a role similar to that of the
two-index epsilon spinors of the two-component spinor algebra; but
it should be evident that in the case of $r$-component
hyperspinors the algebra is more elaborate. In particular, for
$r\geq3$ the epsilon spinor no longer has an interpretation as a
symplectic structure.

Next we introduce the complex matrix space ${\mathbb
C}^{AA'}={\mathbb S}^{A}\otimes{\mathbb S}^{A'}$. An element
$x^{AA'}\in{\mathbb C}^{AA'}$ is \emph{real} if it satisfies the
weak Hermitian property $x^{AA'}= {\bar x}^{A'A}$, where ${\bar
x}^{A'A}$ is the complex conjugate of $x^{AA'}$. We denote the
linear space of real elements of ${\mathbb C}^{AA'}$ by ${\mathbb
R}^{AA'}$. The elements of ${\mathbb R}^{AA'}$ constitute the real
quantum space-time ${\mathcal H}^{N}$ of dimension $N=r^2$. We can
then regard ${\mathbb C}{\mathcal H}^N={\mathbb C}^{AA'}$ as the
complexification of ${\mathcal H}^{N}$.

\section{Chronometric relations on the quantum space-time
${\mathcal H}^{N}$} \label{sec:3}

Consider two points $x^{AA'}$ and $y^{AA'}$ in  ${\mathcal
H}^{N}$, and write $r^{AA'}=x^{AA'} -y^{AA'}$ for the
corresponding separation vector, which is clearly independent of
the choice of origin. In what follows we shall find it useful to
introduce an index-clumping convention (see, e.g., Penrose 1968,
Penrose \& Rindler 1984), and write ${\rm a}=AA'$, ${\rm b}=BB'$,
and so on, according to which a pair of hyperspinor indices, one
primed and the other unprimed, corresponds to a lower case single
vector index. Thus we set $x^{\rm a}=x^{AA'}$, $y^{\rm
a}=y^{AA'}$, $r^{\rm a} = r^{AA'}$, and so on. Then for the
separation vector of the points $x^{\rm a}$ and $y^{\rm a}$ in
${\mathcal H}^{N}$ we write $r^{\rm a}=x^{\rm a}- y^{\rm a}$.

There is a natural causal structure induced on ${\mathcal H}^{N}$
by the so-called \emph{chronometric tensor} $g_{{\rm ab\cdots
c}}$. Making use of the index-clumping convention, we define this
tensor as follows:
\begin{eqnarray}
g_{{\rm ab\cdots c}} = \varepsilon_{AB\cdots C}\,
\varepsilon_{A'B'\cdots C'}.
\end{eqnarray}
The chronometric tensor, which is of rank $r$, is totally
symmetric and is nondegenerate in the sense that for any vector
$r^{\rm a}$ the condition $r^{\rm a}g_{\rm ab\cdots c}=0$ implies
$r^{\rm a}=0$. We shall say that $x^{\rm a}$ and $y^{\rm a}$ in
${\mathbb R}^{AA'}$ are \emph{null separated} if the
\emph{chronometric form} for their separation vanishes:
\begin{eqnarray}
g_{{\rm ab\cdots c}}r^{\rm a}r^{\rm b}\cdots r^{\rm c}=0.
\end{eqnarray}
Null separation is equivalent to the vanishing of the determinant
of the matrix $r^{AA'}$:
\begin{eqnarray}
\varepsilon_{AB\cdots C}\, \varepsilon_{A'B'\cdots C'}
r^{AA'}r^{BB'}\cdots r^{CC'}=0.
\end{eqnarray}

If the hyperspin space has dimension $r=2$, this reduces to the
usual condition for $x^{\rm a}$ and $y^{\rm a}$ to be
null-separated in Minkowski space. For $r>2$, however, the
situation is more complicated on account of the fact that there
are various degrees of nullness that can prevail between two
points. More precisely, when two points of quantum space-time are
null-separated, we shall define the `degree' of nullness by the
rank of the matrix $r^{AA'}$. Null separation of the first degree
is the case for which $r^{AA'}$ is of rank one, and thus satisfies
a system of quadratic relations of the form
\begin{eqnarray}
\varepsilon_{AB\cdots C}\, \varepsilon_{A'B'\cdots C'}
r^{AA'}r^{BB'}=0,
\end{eqnarray}
or equivalently $g_{{\rm ab\cdots c}}r^{\rm a}r^{\rm b}=0$. This
implies in the case of a real separation vector that $r^{AA'}$ can
be expressed in the form
\begin{eqnarray}
r^{AA'}=\pm \alpha^{A}{\bar\alpha}^{A'}
\end{eqnarray}
for some hyperspinor $\alpha^{A}$. If two points have a separation
vector of this form then we say that they are {\it strongly} null
separated. If the sign is positive ({\it resp}. negative), then
$x^{\rm a}$ lies to the future ({\it resp}. past) of $y^{\rm a}$.

In the case of nullness of the second degree, $r^{AA'}$ satisfies
a set of cubic relations given by $g_{{\rm abc\cdots d}} r^{\rm a}
r^{\rm b}r^{\rm c}=0$. In this case $r^{AA'}$ can be put into one
of the following three forms: (a) $r^{AA'}=\alpha^{A}
{\bar\alpha}^{A'}+\beta^A{\bar\beta}^{A'}$, (b) $r^{AA'}= \alpha^A
{\bar\alpha}^{A'}- \beta^{A}{\bar\beta}^{A'}$, and (c) $r^{AA'}=
-\alpha^{A}{\bar\alpha}^{A'}- \beta^{A} {\bar\beta}^{A'}$.

In case (a), $x^{\rm a}$ lies to the future of $y^{\rm a}$, and
$r^{\rm a}$ can be thought of as a `degenerate' future-pointing
time-like vector. In case (c), $x^{\rm a}$ lies to the past of
$y^{\rm a}$, and $r^{\rm a}$ is a degenerate past-pointing
time-like vector. In case (b), $r^{\rm a}$ can be thought of as a
degenerate space-like separation.

A similar analysis can be applied in the case of null separation
of other `intermediate' degrees.

If the determinant of $r^{AA'}$ is nonvanishing, and $r^{AA'}$ is
thus of maximal rank, then the chronometric form is nonvanishing.
In that case $r^{AA'}$ can be represented in the canonical form
\begin{eqnarray}
r^{AA'}=\pm \alpha^{A}{\bar\alpha}^{A'} \pm \beta^{A}
{\bar\beta}^{A'} \pm \cdots \pm \gamma^{A}{\bar\gamma}^{A'},
\label{eq:3.7}
\end{eqnarray}
with the presence of $r$ nonvanishing terms, where the $r$
hyperspinors $\alpha^{A},\beta^{A}, \cdots, \gamma^{A}$ are
linearly independent.

Let us write $(p,q)$ for the numbers of plus and minus signs
appearing in the canonical form for $r^{AA'}$ given in
(\ref{eq:3.7}). We shall call $(p,q)$ the signature of the vector
$r^{AA'}$. The hyperspinors $\alpha^{A}, \beta^{A}, \cdots,
\gamma^{A}$ are determined by $r^{AA'}$ only up to an overall
unitary (or pseudo-unitary) transformation of the form
$\alpha^{A}_n\to U^m_n \alpha^{A}_m$, where $n,m=1,2,\ldots,r$,
and $\alpha^{A}_n = \{\alpha^{A},\beta^{A},\cdots, \gamma^{A}\}$.
The signature $(p,q)$ is, however, an invariant of $r^{AA'}$.

In the cases for which $r^{AA'}$ has signature $(r,0)$ or $(0,r)$
we say that $r^{AA'}$ is time-like future-pointing or time-like
past-pointing, respectively. Then writing
\begin{eqnarray}
\Delta=g_{\rm ab\cdots c}r^{\rm a}r^{\rm b} \cdots r^{\rm c}
\end{eqnarray}
for the associated chronometric form of degree $r$, we define the
\emph{time interval} between the events $x^{\rm a}$ and $y^{\rm
a}$ by the formula
\begin{eqnarray}
\|x-y\| = |\Delta|^{\frac{1}{r}}. \label{eq:3.9}
\end{eqnarray}
It should be evident that in the case $r=2$ we recover the
standard Minkowskian time-interval between the given events.

In summary, the following classification scheme for the separation
between two space-time points can be enunciated. Let $N=r^2$ be
the dimension of the quantum space-time and $(p,q)$ the signature
of the separation vector $r^{\rm a}$. If $p+q=r$ then we say that
the separation is nondegenerate; then if $p=r$, the vector $r^{\rm
a}$ is time-like future-pointing, and if $q=r$, then $r^{\rm a}$
is time-like past-pointing.

If neither $p$ nor $q$ equals $r$, then we say $r^{\rm a}$ is
space-like of type $(p,q)$. Note that in the case of ordinary
Minkowski space, for which $r=2$, the fact that a space-like
vector is necessarily of type $(1,1)$ corresponds to the result
that any space-like vector in four dimensional space-time can be
expressed as the difference between two real null vectors. Any two
such representations for the same space-like vector are related by
a $U(1,1)$ transformation.

On the other hand, if $p+q<r$ then we say that $r^{\rm a}$ is a
\emph{degenerate future-pointing vector} if $q=0$, and a
\emph{degenerate past-pointing vector} if $p=0$, and otherwise a
\emph{degenerate space-like vector}. Clearly, all degenerate
vectors are null in the sense that the corresponding chronometric
form vanishes. If $p=1$ and $q=0$ then $r^{\rm a}$ is
future-pointing and strongly null, and if $p=0$ and $q=1$ then
$r^{\rm a}$ is past-pointing and strongly null. Strong null
separation is the analogue of Minkowskian null separation. The
measure of separation, given by (\ref{eq:3.9}), is nonvanishing if
and only if the separation vector is nondegenerate. The causal
structure of the quantum space-time, however, also brings into
play the various degenerate forms of time-like or null separation.
Thus if $x^{\rm a}$ lies to the future of $y^{\rm a}$ (i.e. if
$x^{\rm a}-y^{\rm a}$ is time-like future-pointing, degenerate
future-pointing, or strongly null future-pointing), and if $y^{\rm
a}$ lies to the future of $z^{\rm a}$, then $x^{\rm a}$ lies to
the future of $z^{\rm a}$.

A striking feature of the causal structure of ${\mathcal H}^{N}$
is that the essential physical features of the causal structure of
Minkowski space are preserved. In particular, the space of future
pointing time-like vectors forms a convex cone, and the convex
hull of this cone includes the future pointing null vectors of all
degrees of degeneracy.

\section{Dynamical trajectories}
\label{sec:4}

Now suppose that the map $\lambda\mapsto x^{AA'}(\lambda)$ defines
a smooth curve $\Gamma$ in ${\mathcal H}^{N}$ for
$\lambda\in[a,b]\subset{\mathbb R}$. Then $\Gamma$ will be said to
be a time-like curve if the tangent vector
\begin{eqnarray}
v^{AA'}(\lambda) = \frac{\rd} {\rd\lambda} x^{AA'}(\lambda)
\end{eqnarray}
is time-like and future-pointing along $\Gamma$. In that case we
define the proper time $s$ elapsed along $\Gamma$ by the integral
\begin{eqnarray}
s = \int_a^b \left[ g_{{\rm ab}\cdots{\rm c}} v^{{\rm a}} v^{\rm
b}\cdots v^{\rm c} \right]^{\frac{1}{r}} \rd \lambda .
\label{eq:3.10}
\end{eqnarray}
For convenience, we can also write (\ref{eq:3.10}) in the
infinitesimal form
\begin{eqnarray}
(\rd s)^r = g_{{\rm ab}\cdots{\rm c}}\rd x^{\rm a} \rd x^{\rm b}
\cdots \rd x^{\rm c}.
\end{eqnarray}
This expression shows that the geometry under consideration here
has a pseudo-Finslerian structure. Finslerian geometries, first
considered by Riemann, and studied extensively by Finsler (see,
e.g., Bao {\it et al}. 2000), have from time to time been proposed
as the basis for generalisations of the theory of relativity. It
is interesting therefore that such a structure arises in a natural
way in the present context. One should note, however, that the
pseudo-Finslerian structures arising in our framework are of a
very particular sort.

For a time-like curve, we can choose the proper time as the
parameter along the curve, in which case the resulting affine
parameterisation of the curve is determined completely up to a
transformation of the form $s\to s+c$ where $c$ is a constant.

The equation of motion for the situation in which $\Gamma$ is a
time-like geodesic is obtained by varying (\ref{eq:3.10}) and
setting the result to zero. As usual, we assume the variation
vanishes at the endpoints; an integration by parts then leads to
the desired result. Writing
\begin{eqnarray}
L=\left( g_{{\rm abc}\cdots{\rm d}} v^{\rm a}v^{\rm b} v^{\rm c}
\cdots v^{\rm d}\right)^{\frac{1}{r}},
\end{eqnarray}
we find that $x^{\rm a}(\lambda)$ describes a geodesic if the
velocity vector $v^{\rm a}$ satisfies the Euler-Lagrange equation
\begin{eqnarray}
\frac{\rd}{\rd\lambda} \left( \frac{\partial L}{\partial v^{\rm
a}} \right) = 0.
\end{eqnarray}
A calculation shows that this condition is given more explicitly
by
\begin{eqnarray}
g_{{\rm abc}\cdots{\rm d}} \frac{\rd v^{\rm b}} {\rd \lambda}
v^{\rm c} \cdots v^{\rm d} = \phi\,g_{{\rm abc}\cdots{\rm d}}
v^{\rm b} v^{\rm c} \cdots v^{\rm d},
\end{eqnarray}
where $\phi=\rd \ln L/\rd\lambda$. If $\lambda$ is chosen to be
proper time, then $\phi=0$ and the geodesic equation takes to form
\begin{eqnarray}
g_{{\rm abc}\cdots{\rm d}} {\dot v}^{\rm b} v^{\rm c} \cdots
v^{\rm d} = 0, \label{eq:3.14}
\end{eqnarray}
where the dot denotes differentiation with respect to the proper
time.

In the case $r=2$ equation (\ref{eq:3.14}) reduces immediately to
the familiar relation ${\dot v}^{\rm a}=0$. To prove that the
geodesic equation (\ref{eq:3.14}) implies ${\dot v}^{\rm a}=0$ in
the case of a quantum space-time for $r\geq2$, it suffices to
examine the case $r=3$. Then we have $g_{{\rm abc}} {\dot v}^{\rm
b} v^{\rm c}=0$, which can be expressed in hyperspinor terms as
\begin{eqnarray}
\epsilon_{ABC}\,\epsilon_{A'B'C'}\,{\dot v}^{BB'} v^{CC'}=0.
\label{eq:3.15}
\end{eqnarray}
This relation in turn can be written
\begin{eqnarray}
{\dot v}^{BB'} v^{CC'}- {\dot v}^{CB'} v^{BC'}- {\dot v}^{BC'}
v^{CB'} + {\dot v}^{CC'} v^{BB'}=0. \label{eq:3.16}
\end{eqnarray}
However, because $\det(v^{AA'})\neq0$ we know that $v^{AA'}$ has
an inverse $u_{AA'}$ satisfying $v^{AA'}u_{BA'}=\delta^A_{\ B}$
and $v^{AA'}u_{AB'}=\delta^{A'}_{\ B'}$. Therefore, contracting
(\ref{eq:3.16}) with $u_{BB'}$ we obtain
\begin{eqnarray}
(u_{BB'}{\dot v}^{BB'})v^{CC'} + (r-2){\dot v}^{CC'} = 0.
\label{eq:3.17}
\end{eqnarray}
This equation shows that if ${\dot v}^{CC'}$ were not zero, then
it would have to be proportional to $v^{CC'}$. However, if that
were the case, then (\ref{eq:3.15}) would imply that
$\det(v^{CC'})=0$, which is contrary to the assumption that
$v^{CC'}$ is time-like. It follows that ${\dot v}^{\rm a}=0$. That
concludes the proof for $r=3$. A similar argument then establishes
for all $r\geq2$ that (\ref{eq:3.14}) implies ${\dot v}^{\rm
a}=0$. Hence we have deduced the following result:

\begin{theorem}
Let $A^{\rm a}$ and $B^{\rm a}$ be quantum space-time points with
the property that $A^{\rm a}-B^{\rm a}$ is time-like and
future-pointing. Then the affinely parametrised geodesic
connecting these points in ${\mathcal H}^{N}$ is the curve
\begin{eqnarray}
X^{\rm a}(s) = B^{\rm a} + \frac{A^{\rm a}-B^{\rm a}}{[\Delta
(A,B)]^{1/r}}\, s
\end{eqnarray}
for $-\infty<s<\infty$, where $\Delta=g_{\rm ab\cdots c}(A^{\rm
a}-B^{\rm a})(A^{\rm b}-B^{\rm b})\cdots (A^{\rm c}-B^{\rm c})$.
\end{theorem}

\section{The Hyper-Poincar\'e group}
\label{sec:5}

The chronometric form $\Delta$ for the separation between two
points in ${\mathcal H}^{N}$ is invariant when the points of
${\mathcal H}^{N}$ are subjected to transformations of the
following type:
\begin{eqnarray}
x^{AA'} \to \lambda^{A}_{B} {\bar\lambda}^{A'}_{B'} x^{BB'} +
b^{AA'} . \label{eq:5.1}
\end{eqnarray}
Here $b^{AA'}$ represents an arbitrary translation in quantum
space-time, $\lambda^{A}_{B}$ is an element of $SL(r,{\mathbb
C})$, and ${\bar\lambda}^{A'}_{B'}$ is the complex conjugate of
$\lambda^A_B$. The relation of this group of transformations to
the Poincar\'e group in the case $r=2$ is evident. Indeed, one of
the attractions of the extension of space-time geometry under
consideration here is that the resulting hyper-Poincar\'e group
admits such a description.

More generally, we observe that the (proper) hyper-Poincar\'e
group preserves the signature of any space-time interval
$r^{AA'}$, whether or not the interval is degenerate, and hence
leaves the causal relations between events unchanged.

We refer to a transformation of the form $r^{\rm a}\to L^{\rm
a}_{\rm b}r^{\rm b}$ as a hyper-Lorentz transformation if $L^{\rm
a}_{\rm b}= \lambda^{A}_{B} {\bar\lambda}^{A'}_{B'}$ for some
element $\lambda^{A}_{B} \in SL(r,{\mathbb C})$. The (real)
dimension of the hyper-Lorentz group is $2r^2-2$, and the
dimension of the hyper-Poincar\'e group is thus $3r^2-2$. It is
interesting to observe that the dimension of the hyper-Poincar\'e
group grows linearly with the dimension of the quantum space-time
itself, which is given by $r^2$. This can be contrasted with the
dimension of the group arising if we endow ${\mathbb R}^{N}$ with
a standard Lorentzian metric with signature $(1,r^2-1)$. In that
case the pseudo-orthogonal group has real dimension
$\frac{1}{2}r^2(r^2-1)$, which together with the translation group
gives a total dimension of $\frac{1}{2}r^2(r^2+1)$. The
parsimonious dimensionality of the hyper-Poincar\'e group is due
to the fact that it preserves a rather delicate system of causal
relations holding between pairs of points in quantum space-time.

\section{Symmetries and conservation laws}
\label{sec:6}

In Minkowski space the symmetries of the Poincar\'e group are
associated with a ten-parameter family of Killing vectors. That is
to say, for $r=2$ we have the Minkowski metric $g_{\rm ab} =
\varepsilon_{AB}\varepsilon_{A'B'}$, and the Poincar\'e group is
generated by the ten-parameter family of vector fields $\xi^{\rm
a}(x)$ on ${\mathfrak M}^4$ satisfying ${\mathcal L}_\xi g_{\rm
ab}=0$, where ${\mathcal L}_\xi$ denotes the Lie derivative with
respect to $\xi^{\rm a}(x)$. Now for any vector field $\xi^{\rm
a}$ and any symmetric tensor field $g_{\rm ab}$ we have
\begin{eqnarray}
{\mathcal L}_\xi g_{\rm ab} = \xi^{\rm c} \nabla_{\!\rm c}g_{\rm
ab} + 2g_{{\rm c}({\rm a}} \nabla_{\!{\rm b})} \xi^{\rm c} .
\end{eqnarray}
If $g_{\rm ab}$ is the metric and $\nabla_{\!\rm a}$ denotes the
associated Christoffel derivative satisfying $\nabla_{\!\rm
a}g_{\rm bc}=0$, we obtain $\nabla_{\!({\rm a}}\xi_{{\rm b})}=0$,
where $\xi_{\rm a}=g_{\rm ab}\xi^{\rm b}$. The condition
${\mathcal L}_\xi g_{\rm ab}=0$ therefore implies that $\xi^{\rm
a}$ is a Killing vector.

We have taken the trouble to spell out the case for $r=2$ in order
to highlight the contrast with the situation for general $r$.
Clearly for $r>2$ we have no Riemannian metric, and the usual
relation between symmetries and Killing vectors is lost. What
survives, however, is of considerable interest. More specifically,
to generate a symmetry of the quantum space-time ${\mathcal
H}^{N}$ the vector field $\xi^{\rm a}$ has to satisfy
\begin{eqnarray}
{\mathcal L}_\xi g_{\rm ab\cdots c} = 0,
\end{eqnarray}
where $g_{\rm ab\cdots c}$ is the chronometric tensor. Now for a
general vector field $\xi^{\rm a}$ and a general symmetric tensor
$g_{\rm ab\cdots c}(x)$ we have
\begin{eqnarray}
{\mathcal L}_\xi g_{\rm ab\cdots c} = \xi^{\rm d} \nabla_{\!\rm d}
g_{\rm ab\cdots c} + r\,g_{{\rm d}({\rm a\cdots b}} \nabla_{\!{\rm
c})} \xi^{\rm d} .
\end{eqnarray}
In the case of a quantum space-time we let $\nabla_{\!{\rm a}}$ be
the flat connection for which $\nabla_{\!{\rm a}} g_{\rm bc\cdots
d}=0$. Then to generate a symmetry of the chronometric structure
of ${\mathcal H}^{N}$ the vector field $\xi^{\rm a}$ must satisfy
\begin{eqnarray}
g_{{\rm d}({\rm a\cdots b}} \nabla_{\!{\rm c})} \xi^{\rm d} = 0,
\label{eq:4.5}
\end{eqnarray}
which serves as the analogue of Killing's equation. Equation
(\ref{eq:4.5}) can be written in a more suggestive form if we
define a symmetric tensor $\xi_{\rm ab\cdots c}$ of rank $r-1$ by
\begin{eqnarray}
\xi_{\rm ab\cdots c} = g_{\rm ab\cdots cd}\xi^{\rm d} .
\label{eq:4.6}
\end{eqnarray}
Then (\ref{eq:4.5}) says that $\xi_{\rm ab\cdots c}$ satisfies the
conditions for a symmetric \emph{Killing tensor}:
\begin{eqnarray}
\nabla_{\!({\rm a}}\xi_{\rm bc\cdots d)} = 0 . \label{eq:4.7}
\end{eqnarray}
Thus we see that ${\mathcal H}^{N}$ provides an example of a
symmetry group generated by a family of Killing tensors.

\begin{theorem}
The symmetries of the quantum space-time ${\mathcal H}^{N}$ are
generated by a system of $3r^2-2$ irreducible symmetric Killing
tensors of rank $r-1$.
\end{theorem}

The significance of Killing tensors is that they are associated
with the existence of conserved quantities. A well-known example
of a conserved quantity associated with an irreducible Killing
tensor (that is to say, a Killing tensor that cannot be expressed
as a sum of products of Killing vectors) is Carter's fourth
integral of the equations of motion for geodesics and
charged-particle orbits in the Kerr and Kerr-Newman solutions of
Einstein's equations (Carter 1968, Walker \& Penrose 1970,
Hughston, {\it et al}. 1971, Hughston \& Sommers 1973, Penrose \&
Rindler 1986).

In the present context it follows that if the vector field $v^{\rm
a}(x)$ satisfies the geodesic equation, which on a chronometric
space of dimension $r^2$ is given by
\begin{eqnarray}
g_{\rm abc\cdots d}\left( v^{\rm e}\nabla_{\!\rm e}v^{\rm
b}\right) v^{\rm c}\cdots v^{\rm d} = 0,
\end{eqnarray}
and if $\xi_{\rm ab\cdots c}$ is the Killing tensor of rank $r-1$
given by (\ref{eq:4.6}), then we have the conservation law
\begin{eqnarray}
v^{\rm e}\nabla_{\!\rm e}\left( \xi_{\rm ab\cdots c} v^{\rm a}
v^{\rm b}\cdots v^{\rm c}\right) = 0.
\end{eqnarray}
In other words, $g_{\rm ab\cdots cd} v^{\rm a}v^{\rm b}\cdots
v^{\rm c}\xi^{\rm d}$ is a constant of the motion.

Thus in higher-dimensional quantum space-times the apparatus of
conservation laws and symmetry principles remains intact in the
absence of a pseudo-Riemannian metric. In particular, the
conservation of hyper-relativistic momentum and angular momentum
for a system of interacting particles can be given a well-defined
formulation, the basic principles of which are similar to those
applicable in the Minkowskian case. For this purpose it is useful
to introduce the notion of an `elementary system' or particle in
hyper-relativistic mechanics. Such a system is defined by its
hyper-relativistic momentum and angular momentum.

The hyper-relativistic momentum of an elementary system is given
by a momentum covector $P_{\rm a}$. The associated mass $m$ is
then defined by the invariant
\begin{eqnarray}
m=\left( g^{\rm ab\cdots c}P_{\rm a}P_{\rm b}\cdots P_{\rm c}
\right)^{\frac{1}{r}}.
\end{eqnarray}
The hyper-relativistic angular momentum of an elementary system is
given by a tensor $L^{\rm b}_{\rm a}$ of the form
\begin{eqnarray}
L^{\rm b}_{\rm a} = l^{B}_{A}\delta^{B'}_{A'} + {\bar l}^{B'}_{A'}
\delta^{B}_{A} ,
\end{eqnarray}
where the hyperspinor $l^{B}_{A}$ is required to be trace-free:
$l^{A}_{A}=0$. The angular momentum is defined with respect to a
choice of origin, in such a manner that under a change of origin
defined by a shift vector $b^{\rm a}$ we have $l^{B}_{A}\to l^B_A
+ P_{AC'}b^{BC'}$. In the case $r=2$ these formulae reduce to the
usual formulae for relativistic momentum and angular momentum in a
Minkowskian setting. The real covector
\begin{eqnarray}
S_{AA'} = \ri m^{-1} \left( l^B_A P_{A'B} - {\bar l}^{B'}_{A'}
P_{AB'} \right)
\end{eqnarray}
is invariant under a change of origin, and carries the
interpretation of the intrinsic spin of the elementary system. The
magnitude $S$ of the spin is then defined by $S=|g^{\rm ab\cdots
c} S_{\rm a}S_{\rm b}\cdots S_{\rm c}|^{\frac{1}{r}}$.

In the case of a set of interacting hyper-relativistic systems we
require that the total momentum and angular momentum should both
be conserved. This then implies conservation of the total mass and
spin.

\section{Geometry of complex null-separation}
\label{sec:7}

In Minkowski space it is useful to examine the geometry of the
space of complex null vectors at a point in the space-time. Thus
we take the case $r=2$, and consider complex vectors $z^{\rm a}$
satisfying $g_{\rm ab}z^{\rm a}z^{\rm b}=0$. In spinor terms this
implies that $z^{\rm a}$ can be written in the form
\begin{eqnarray}
z^{{\bf AA}'} = \alpha^{\bf A}\beta^{{\bf A}'}. \label{eq:7.1}
\end{eqnarray}
In we take a projective point of view, then up to overall scale
the space of complex vectors at a point in Minkowski space can be
regarded as a complex projective space ${\mathbb P}^3$. The null
directions constitute a quadric ${\mathbb Q}^2$ in that space,
which owing to the decomposition (\ref{eq:7.1}) has the structure
of a doubly ruled surface ${\mathbb Q}^2={\mathbb P}^1 \times
{\mathbb P}^1$. We can identify the first set of lines (the
$\alpha$-lines) with the projective unprimed spinors, and the
second set of lines (the $\beta$-lines) with the projective primed
spinors. The quadric ${\mathbb Q}^2$ is ruled in such a manner
that two lines of the same type do not intersect, whereas two
lines of the opposite type intersect at a point in ${\mathbb
Q}^2$---i.e. the null direction they together determine.

In the case of a general quantum space-time, we consider the space
of complex vectors at a point of ${\mathcal H}^{N}$, and examine
the corresponding space of directions, which has the structure of
a complex projective space ${\mathbb P}^{r^2-1}$. The vanishing of
the chronometric form $g_{\rm ab \cdots c}z^{\rm a}z^{\rm b}\cdots
z^{\rm c}$ identifies the space of complex null directions as a
hypersurface of degree $r$ in ${\mathbb P}^{r^2-1}$ which we shall
call ${\mathfrak N}^{r^2-2}$.

The points of ${\mathfrak N}^{r^2-2}$ correspond to `weakly' null
directions. The strongly null directions in ${\mathfrak
N}^{r^2-2}$, corresponding to those for which the associated null
vectors are of minimal rank and hence of the form
$z^{AA'}=\alpha^A\beta^{A'}$, constitute a subvariety ${\mathbb
Q}^{2r-2}\subset{\mathfrak N}^{r^2-2}$ defined by the mutual
intersection of a system of quadrics, given by the equation
$g_{\rm ab \cdots c}z^{\rm a} z^{\rm b}=0$. In this case we have
${\mathbb Q}^{2r-2}={\mathbb P}^{r-1} \times {\mathbb P}^{r-1}$,
and we can identify the two systems of $(r-1)$-planes by which
${\mathbb Q}^{2r-2}$ is foliated, which we refer to as
$\alpha$-planes and $\beta$-planes, as the spaces of projective
unprimed and primed hyperspinors, respectively.

The various null directions of intermediate degree correspond to
points in ${\mathfrak N}^{r^2-2}$ lying on the linear spaces
spanned by the joins of $k$ points in ${\mathbb Q}^{2r-2}$
($k=2,3,\ldots,r$). The degree of nullness, as defined in
\S\ref{sec:3}, is given by the integer $d=k-1$.

In the case $r=3$, for example, the space of directions at a point
in ${\mathcal H}^9$ is ${\mathbb P}^{8}$, and the null directions
constitute a cubic hypersurface ${\mathfrak N}^{7}\subset {\mathbb
P}^8$. The null directions of the first degree (i.e. the totally
null directions) lie in the doubly foliated surface ${\mathbb Q}^4
={\mathbb P}^{2} \times {\mathbb P}^{2}$ in ${\mathfrak N}^{7}$.
The points of ${\mathfrak N}^{7}$ all lie on the `first join' of
${\mathbb Q}^4$ with itself; in other words, any point of
${\mathfrak N}^7$ lies on a line joining two points of ${\mathbb
Q}^4$. The space ${\mathfrak N}^7\!\setminus\!{\mathbb Q}^4$ then
consists of null directions that are strictly of the second
degree. (A null direction is strictly of the second degree if it
is of the second degree but not also of the first degree.) Note
that any point of ${\mathbb P}^8$ can be represented as the join
of three points in ${\mathbb Q}^4$.

In the case $r=4$, the space of directions at a point in
${\mathcal H}^{16}$ is ${\mathbb P}^{15}$, and the null directions
constitute a quadric hypersurface ${\mathfrak N}^{14}\subset
{\mathbb P}^{15}$. The null directions of the first degree lie on
the doubly foliated surface ${\mathfrak N}^{6}_{(1)}={\mathbb
P}^{3} \times {\mathbb P}^{3}$ in ${\mathfrak N}^{14}$. The null
directions of the second degree lie on the first join of ${\mathbb
Q}^6$ with itself: ${\mathfrak N}_{(2)}=J_1({\mathbb Q}^{6})$. The
null directions of the third degree lie in ${\mathfrak N}_{(3)}
=J_2({\mathbb Q}^{6})$ and constitute the general elements of
${\mathfrak N}^{14}$.

It is interesting to note a distinction between the Minkowskian
case $r=2$ and higher dimensional quantum space-times. In
Minkowski space, the space of complex null directions at a given
point corresponds to a nondegenerate quadric ${\mathbb Q}^2$ in
${\mathbb P}^3$, which is doubly ruled in the sense that ${\mathbb
Q}^2={\mathbb P}^1\times{\mathbb P}^1$. In fact, any nondegenerate
quadric in ${\mathbb P}^3$ has this structure, and by an
automorphism of ${\mathbb P}^3$ one can transform any such quadric
into another.

For $r>2$, however, this is generally not the case. For example,
in the case $r=3$ the general cubic hypersurface in ${\mathbb
P}^8$ does not contain within it a doubly foliated subvariety
${\mathbb Q}^4={\mathbb P}^2\times{\mathbb P}^2$. The space of
cubic hypersurfaces in ${\mathbb P}^8$ has (complex) dimension
$164$. If we factor out by the group of projective automorphisms
of ${\mathbb P}^8$, which is of dimension $80$, then we are left
with the $84$-dimensional moduli-space for cubic forms in
${\mathbb P}^8$. The chronometric form of our quantum space-time
geometry represents a single point in this moduli-space, and as
such constitutes a special cubic surface in ${\mathbb P}^8$. Such
a hypersurface is in fact completely characterised by the
existence of an embedded hyperspinorial subvariety of the form
${\mathbb Q}^4={\mathbb P}^2\times{\mathbb P}^2$. Any two cubic
hypersurface in ${\mathbb P}^8$ admitting a hyperspinorial
subvariety can be transformed into one another by an automorphism
of ${\mathbb P}^8$, and thus represent the same point in the
moduli-space. A similar observation applies for all $r>2$.

\section{Generalised Klein representation}
\label{sec:8}

To proceed further it will be useful if we set the foregoing
material in a geometric context that emphasises the conformal
properties of the chronometric form. To this end we let ${\mathbb
T}^{\boldsymbol\alpha}$ denote the complex vector space of
dimension $2r$ given by the pair $({\mathbb S}^{A},{\mathbb
S}_{A'})$. Let us write $Z^{\alpha}= (\omega^{A},\pi_{A'})$ for a
typical element of ${\mathbb T}^{{\alpha}}$. Such an element will
be referred to as a hypertwistor. For a brief introduction to the
theory of hypertwistors (also called `generalised twistors') see
Hughston (1978). Let ${\mathbb T}_{{\alpha}}=({\mathbb
S}_{A},{\mathbb S}^{A'})$ denote the space of dual hypertwistors.
A natural pseudo-Hermitian structure can be introduced on the
geometry of hypertwistors by means of the complex conjugation
operation that maps $(\omega^{A},\pi_{A'}) \in{\mathbb
T}^{{\alpha}}$ to $({\bar\pi}_{A},{\bar\omega}^{A'}) \in{\mathbb
T}_{{\alpha}}$. The corresponding pseudo-Hermitian form is then
given by
\begin{eqnarray}
Z^{{\alpha}} {\bar Z}_{{\alpha}} = \omega^{A} {\bar\pi}_{A} +
\pi_{A'} {\bar\omega}^{A'},
\end{eqnarray}
and it is straightforward exercise to verify that the inner
product $Z^{\alpha}{\bar Z}_{{\alpha}}$ is invariant under the
action of the group $U(r,r)$.

The space ${\mathbb P}^{2r-1}$ of projective hypertwistors is a
natural starting point for analysing the conformal geometry of
complex quantum space-time, which can be regarded as the
Grassmannian variety ${\mathbb V}^{r^2}$ of projective
$(r-1)$-planes in ${\mathbb P}^{2r-1}$. More precisely, ${\mathbb
V}^{r^2}$ can be understood as the complex quantum space-time
${\mathbb C}{\mathcal H}^{r^2}$ introduced earlier, together with
some structure added at infinity. Thus ${\mathbb V}^{r^2}$ is to
be understood as a compactification of ${\mathbb C}{\mathcal
H}^{r^2}$. The `finite' points of ${\mathbb V}^{r^2}$ correspond
to the linear subspaces of ${\mathbb P}^{2r-1}$ that are
determined by a relation of the form
\begin{eqnarray}
\omega^{A} = \ri x^{AA'} \pi_{A'}
\end{eqnarray}
for fixed $x^{AA'}$. The aggregate of such $(r-1)$-planes
constitute the points of ${\mathbb C}{\mathcal H}^{r^2}$. The
$(r-1)$-planes for which $x^{AA'}$ is Hermitian then constitute
the points of the real space ${\mathcal H}^{r^2}$.

The conformal structure of quantum space-time is implicit in the
various possibilities arising for the intersections of
$(r-1)$-planes in hypertwistor space. A pair of $(r-1)$-planes in
${\mathbb P}^{2r-1}$ in general will not intersect. This general
lack of intersection corresponds to the nonvanishing of the
chronometric form for the corresponding quantum space-time points.
In this connection we note that the chronometric form $\Delta=
g_{{\rm ab\cdots c}}r^{\rm a}r^{\rm b}\cdots r^{\rm c}$ introduced
earlier for the pairs of real quantum space-time points is also
well-defined for pairs of complex quantum space-time points. Now
an $(r-1)$-plane in ${\mathbb P}^{2r-1}$ is represented by a
simple skew hypertwistor $P^{{\alpha}{\beta}\cdots{\gamma}}$ of
rank $r$. If $r=2$, we recover the fact that ordinary space-time
points correspond to projective lines in ${\mathbb P}^{3}$, which
in turn correspond to simple antisymmetric twistors of rank two.
By a \emph{simple} skew hypertwistor we mean one of the form
\begin{eqnarray}
P^{{\alpha}{\beta}\cdots{\gamma}} = A^{[{\alpha}}B^{{\beta}}\cdots
C^{{\gamma}]}
\end{eqnarray}
for some collection $A^{\alpha},B^{\alpha},\cdots,C^{\alpha}$ of
$r$ hypertwistors (all of which lie on the given plane). Suppose
that the simple skew hypertwistors
$P^{{\alpha}{\beta}\cdots{\gamma}}$ and
$Q^{{\alpha}{\beta}\cdots{\gamma}}$ represent, respectively, the
$(r-1)$-planes $P$ and $Q$ in ${\mathbb P}^{2r-1}$. Then a
necessary and sufficient condition for the vanishing of the
chronometric form for the corresponding quantum space-time points
is
\begin{eqnarray}
\varepsilon_{\alpha\beta\cdots\gamma\rho\sigma\cdots\tau}
P^{\alpha\beta\cdots\gamma} Q^{\rho\sigma\cdots\tau} = 0,
\label{eq:5.4}
\end{eqnarray}
where
$\varepsilon_{{\alpha}{\beta}\cdots{\gamma}{\rho}{\sigma}\cdots{\tau}}$
is the totally skew hypertwistor of rank $2r$. We note that
(\ref{eq:5.4}) is symmetric ({\it resp}., antisymmetric) under the
interchange of $P^{{\alpha}{\beta}\cdots{\gamma}}$ and
$Q^{{\alpha}{\beta}\cdots{\gamma}}$ if $r$ is even ({\it resp}.,
odd). The vanishing of the form (\ref{eq:5.4}) is the condition
that the projective planes $P$ and $Q$ contain a point in common.
Equivalently, this means that the skew hypertwistors
$P^{{\alpha}{\beta}\cdots{\gamma}}$ and
$Q^{{\rho}{\sigma}\cdots{\tau}}$ contain at least one hypertwistor
as a common factor. Thus we have deduced the following result.

\begin{proposition}
A necessary and sufficient condition for a pair of quantum
space-time points to be weakly null separated is that the
corresponding $(r-1)$-planes in ${\mathbb P}^{2r-1}$ should
intersect.
\end{proposition}

More generally, the degree $d$ of null separation for a pair of
quantum space-time events is given by $d=r-m-1$, where $m$ is the
dimensionality of the intersection of the corresponding
$(r-1)$-planes in ${\mathbb P}^{2r-1}$. The possible degrees of
null separation are given by $d=1,2,\ldots,r-1$. If we interpret
(as usual) the case of no intersection as an intersection of
dimension $-1$, then a non-null separation between the
corresponding quantum space-time points can be interpreted as a
`separation of degree $r$'. Thus separations of degree less than
$r$ are all null, whereas a separation of degree $r$ is non-null.
The degree of separation of a pair of points, we recall, is given
by the rank of the separation matrix $r^{AA'}=x^{AA'}-y^{AA'}$.
Equivalently, given two skew hypertwistors $P^{{\alpha}{\beta}
\cdots{\gamma}}$ and $Q^{{\alpha}{\beta} \cdots{\gamma}}$, each
with $r$ indices, let us form the dual hypertwistor by
$Q_{{\alpha}{\beta}\cdots{\gamma}} =\varepsilon_{{\alpha}{\beta}
\cdots {\gamma}{\rho}{\sigma} \cdots{\tau}}Q^{{\rho}
{\sigma}\cdots{\tau}}$. Then $d$ is given by the maximum number of
index contractions we can make between $P^{{\alpha}{\beta}
\cdots{\gamma}}$ and $Q_{{\alpha}{\beta}\cdots{\gamma}}$ without
obtaining the result zero. If a single index contraction gives
zero, this corresponds to the case where $P^{{\alpha}{\beta}
\cdots{\gamma}}$ is proportional to $Q^{{\alpha}{\beta}
\cdots{\gamma}}$. Thus $d=0$ (separation of degree zero) can be
interpreted as the `degenerate' case where the two quantum
space-time points coincide.

\section{Quantum infinity}
\label{sec:9}

As indicated in the previous section, for any skew hypertwistor
$P^{{\alpha}{\beta}\cdots{\gamma}}$  of rank $r$ we define its
dual $P_{{\alpha}{\beta}\cdots{\gamma}}$ by the relation
\begin{eqnarray}
P_{{\alpha}{\beta}\cdots{\gamma}} =
\varepsilon_{{\alpha}{\beta}\cdots{\gamma}
{\rho}{\sigma}\cdots{\tau}} P^{{\rho}{\sigma}\cdots{\tau}} .
\end{eqnarray}
Here $\varepsilon_{{\alpha}{\beta}\cdots{\gamma}
{\rho}{\sigma}\cdots{\tau}}$ is the totally skew hypertwistor of
rank $2r$, which is unique up to an overall scale factor.
Depending on whether $r$ is even or odd, we have the following
interchange relations:
\begin{eqnarray}
\varepsilon_{{\alpha}{\beta}\cdots{\gamma}{\rho}{\sigma}\cdots
{\tau}} = \pm\varepsilon_{{\rho}{\sigma} \cdots{\tau} {\alpha}
{\beta} \cdots {\gamma}}\, .
\end{eqnarray}
Thus if $r$ is even, then once the scale of the totally skew
hypertwistor is fixed we obtain a \emph{symmetric inner product}
on the space of skew hypertwistors of rank $r$, which we can
denote symbolically by
\begin{eqnarray}
\langle P,Q\rangle =
\varepsilon_{{\alpha}{\beta}\cdots{\gamma}{\rho}{\sigma}\cdots{\tau}}\,
P^{{\alpha}{\beta}\cdots{\gamma}} Q^{{\rho}{\sigma}\cdots{\tau}} .
\label{eq:6.3}
\end{eqnarray}
On the other hand if $r$ is odd then the inner product
(\ref{eq:6.3}) is a symplectic structure. In this respect the
cases for even $r$ and odd $r$ are quite distinct. We shall return
to this issue later when we specialise to the symmetric case.

Recall that under complex conjugation the skew hypertwistor
$P^{{\alpha}{\beta}\cdots{\gamma}}$ becomes ${\bar P}_{{\alpha}
{\beta}\cdots{\gamma}}$. If $P^{{\alpha}{\beta}\cdots{\gamma}}$ is
simple, thus corresponding to an $(r-1)$-plane $P$ in ${\mathbb
P}^{2r-1}$, then we say that $P$ is a \emph{real} plane if ${\bar
P}_{{\alpha}{\beta}\cdots{\gamma}}$ is proportional to
$P_{{\alpha}{\beta}\cdots{\gamma}}$. The real $(r-1)$-planes of
${\mathbb P}^{2r-1}$ thus defined correspond to the real points of
quantum space-time.

The points at infinity in the compactified quantum space-time
${\mathcal V}^{r^2}={\mathbb C}{\mathcal H}_{\sharp}^{\,r^2}$ can
be described as follows. In the hypertwistor space ${\mathbb
P}^{2r-1}$ we choose a real $(r-1)$-plane $I$ represented by a
simple skew hypertwistor $I^{{\alpha}{\beta}\cdots{\gamma}}$. The
point $I$ in ${\mathcal V}^{r^2}$ corresponding to the
$(r-1)$-plane $I$ in ${\mathbb P}^{2r-1}$ will be called the
\emph{point at infinity}. The cone in ${\mathcal V}^{r^2}$
consisting of all points that are chronometrically null-separated
from $I$ will be called \emph{null infinity}. (There will be no
ambiguity if we use the symbol $I$ to denote both the point $I$ in
${\mathcal V}^{r^2}$ and the corresponding $(r-1)$-plane in
${\mathbb P}^{2r-1}$.) It should be evident that null infinity has
a rich structure, with various domains that can be classified
according to their degree of null separation from the point $I$.

The `finite' points of ${\mathcal V}^{r^2}$ are those for which
the chronometric separation from $I$ is non-null, i.e. those
points $P$ for which $\langle P,I\rangle\neq0$ with respect to the
inner product (\ref{eq:6.3}).

\begin{proposition}
In the case of two finite quantum space-time points the
chronometric form $\Delta$ is given by the following ratio:
\begin{eqnarray}
\Delta(P,Q) = \frac{
\varepsilon_{{\alpha}{\beta}\cdots{\gamma}{\rho}{\sigma}\cdots{\tau}}
P^{{\alpha}{\beta}\cdots{\gamma}} Q^{{\rho}{\sigma}\cdots{\tau}}}
{(\varepsilon_{{\alpha}{\beta}\cdots{\gamma}{\rho}{\sigma}\cdots{\tau}}
P^{{\alpha}{\beta}\cdots{\gamma}} I^{{\rho}{\sigma}\cdots{\tau}})
(\varepsilon_{{\alpha}{\beta}\cdots{\gamma}{\rho}{\sigma}\cdots{\tau}}
Q^{{\alpha}{\beta}\cdots{\gamma}} I^{{\rho}{\sigma}\cdots{\tau}})}
. \label{eq:6.4}
\end{eqnarray}
\end{proposition}

Equivalently we can write $\Delta=\langle P,Q\rangle/\langle
P,I\rangle \langle Q,I\rangle$. If $P$ and $I$ are not null
separated, then we can choose the scales of
$P^{{\alpha}{\beta}\cdots{\gamma}}$ and
$I^{{\alpha}{\beta}\cdots{\gamma}}$ such that $\langle P, I\rangle
=1$, without loss of generality, and similarly for
$Q^{{\alpha}{\beta}\cdots{\gamma}}$ and
$I^{{\alpha}{\beta}\cdots{\gamma}}$. This leads to a further
simplification in formula (\ref{eq:6.4}).

In general, even in the absence of such a simplification, we note
that $\Delta(P,Q)$ is independent of the scale of
$P^{{\alpha}{\beta}\cdots{\gamma}}$ and
$Q^{{\alpha}{\beta}\cdots{\gamma}}$. On the other hand,
$\Delta(P,Q)$ does depend on the scale of
$\varepsilon_{{\alpha}{\beta}\cdots{\gamma}{\rho}{\sigma}\cdots{\tau}}$
and the scale of $I^{{\alpha}{\beta}\cdots{\gamma}}$. It has an
epsilon `weight' of $-1$ and an $I$ `weight' of $-2$ (cf. Hughston
\& Hurd 1982). However, if we form the ratio associated with four
hypertwistors $P$, $Q$, $R$, and $S$, given by
\begin{eqnarray}
\frac{\Delta(P,Q)}{\Delta(R,S)} = \frac{\|p-q\|^r}{\|r-s\|^r},
\label{eq:6.5}
\end{eqnarray}
where $p$, $q$, $r$, and $s$ are the quantum space-time points
corresponding to $P$, $Q$, $R$, and $S$, respectively, then we
obtain an expression that is absolute---that is to say, a
geometric invariant. This is because $\Delta(P,Q)$ has the
`dimensionality' of time raised to the power $r$; whereas the
ratio (\ref{eq:6.5}) arises as a comparison of two such time
intervals, and thus is dimensionless. The basic chronometric
geometry, with infinity chosen as indicated above, admits no
absolute or `preferred' unit of time: in this geometry only ratios
of time intervals have an absolute meaning.

\section{Cosmological infinity}
\label{sec:10}

There is, however, no reason {\it a priori} why such a `minimal'
structure should prevail at infinity. Other choices are in
principle available for $I^{{\alpha}{\beta}\cdots{\gamma}}$, and
these have the effect of giving ${\mathcal V}^{r^2}$ the structure
of a cosmological model. In the case $r=2$, for example, if
$I^{\alpha\beta}$ is chosen to be real and non-simple, then the
quadratic form $\varepsilon_{\alpha\beta \gamma\delta}
I^{\alpha\beta} I^{\gamma\delta}$, which has an epsilon weight of
one and an $I$-weight of two, has the dimensionality of inverse
squared-time. Hence in this case there \emph{is} a preferred unit
of time.

To pursue this point further, we recall that ${\mathcal V}^4$ has
the structure of a quadric $\Omega$ in ${\mathbb P}^5$. More
specifically, for $r=2$ the space of skew rank two twistors is
${\mathbb C}^6$, which is projectively ${\mathbb P}^5$, and
${\mathcal V}^4$ is the locus defined by the homogeneous quadratic
equation $\varepsilon_{\alpha\beta\gamma\delta} X^{\alpha\beta}
X^{\gamma\delta}=0$. Infinity in ${\mathcal V}^4$ can then be
defined by the intersection of ${\mathcal V}^4$ in ${\mathbb P}^5$
with the projective 4-plane $I$ given by the equation
$\varepsilon_{\alpha\beta\gamma\delta} I^{\alpha\beta}
X^{\gamma\delta}=0$. If $I^{\alpha\beta}$ is simple, then $I^4$ is
tangent to ${\mathcal V}^4$, and the intersection is a cone---the
null cone at infinity. On the other hand, if $I^{\alpha\beta}$ is
not simple, then the intersection $I^4\cap{\mathcal V}^4$ is a
3-quadric. The resulting geometry, if $I^{\alpha\beta}$ is real,
is that of de~Sitter space, and the parameter $\lambda=\left(
\varepsilon_{\alpha\beta\gamma\delta} I^{\alpha\beta}
I^{\gamma\delta}\right)^2$ has the interpretation of being the
associated \emph{cosmological constant}. The de Sitter group then
consists of those projective transformations of ${\mathbb P}^5$
that preserve both $\Omega$ and the point $I$. In fact, with the
incorporation of some additional structure at infinity, the entire
class of Robertson-Walker cosmological models can be represented
in this way (Penrose \& Rindler 1984, Hurd 1985, 1995, Penrose
1995).

For general $r$ a similar situation arises---in other words, the
choice of structure at infinity gives rise in general to a
chronometric metric that is not flat, thus giving ${\mathcal
V}^{r^2}$ the character of a cosmological model. The key point is
that, whereas in the case of a standard four-dimensional
cosmological model based on Einstein's theory the existence of
structure at infinity has a bearing on the geometry of space-time
alone, in the case of a hypercosmology the structure at infinity
also has implications for microscopic physics. In particular,
whereas in the four-dimensional de~Sitter cosmology the relevant
structure at infinity contains the information of a single
dimensional constant (the cosmological constant), in the
higher-dimensional situation there will in general be a number of
such dimensional constants emerging as geometrical invariants of
the theory. Thus within the same overall geometric framework one
has the scope for introducing structure (or what amounts to the
same thing---the breaking of symmetry) both on a global or
cosmological scale, as well as on those scales of distance, time,
and energy associated with the phenomenology of elementary
particles.

In order to prepare the groundwork necessary as a basis for
investigating this idea in more detail we must now introduce the
particular structure in ${\mathcal V}^{r^2}$ needed to make its
relation to ordinary four-dimensional space-time apparent.

\section{Segr\'e embedding and symmetry breaking mechanism}
\label{sec:11}

Let us therefore consider the mechanisms for symmetry breaking at
our disposal in the case of a standard `flat' quantum space-time
endowed with a canonical reality structure and null infinity. We
shall demonstrate that the breaking of symmetry in quantum
space-time is intimately linked to the notion of quantum
entanglement.

In practical terms the breaking of symmetry can be represented by
an `index decomposition'. The point is that if the dimension $r$
of the hyperspin space is not a prime number, then a natural
method of breaking the symmetry arises by consideration of the
decomposition of $r$ into factors. The specific assumption that we
shall make at this juncture will be that the dimension of the
hyperspin space ${\mathbb S}^{A}$ is \emph{even}. Then we write
$r=2n$, where $n=1,2,\ldots$, and set ${\mathbb S}^{A}={\mathbb
S}^{{\bf A}i}$, where ${\bf A}$ is a standard two-component spinor
index, and $i$ will be called an \emph{internal} index $(i=1,2,
\ldots,n)$. Thus we can write ${\mathbb S}^{{\bf A}i}={\mathbb
S}^{{\bf A}} \otimes{\mathbb H}^{i}$, where ${\mathbb S}^{{\bf
A}}$ is a standard spin space of dimension two, and ${\mathbb
H}^{i}$ is a complex vector space of dimension $n$. The breaking
of the symmetry then amounts to the fact that we can identify the
hyperspin space with the tensor product of these two spaces.

We shall assume, moreover, that as far as the internal space is
concerned, there is a canonical anti-linear isomorphism between
the complex conjugate of the internal space ${\mathbb H}^{i}$ and
the dual space ${\mathbb H}_{i}$. In other words, if
$\psi^i\in{\mathbb H}^{i}$, then we can write ${\bar\psi}_i$ for
the complex conjugate of $\psi^i$, where ${\bar\psi}_i\in{\mathbb
H}_{i}$. Therefore, ${\mathbb H}^{i}$ is a complex Hilbert
space---and indeed although for the moment we consider for
technical simplicity the case for which $n$ is finite, one should
have in mind also the infinite dimensional situation.

For the other hyperspin spaces we write ${\mathbb S}_{A}={\mathbb
S}_{{\bf A}i}$, ${\mathbb S}^{A'}={\mathbb S}^{{\bf A}'}_{\ \ i}$,
and ${\mathbb S}_{A'}={\mathbb S}_{{\bf A}'}^{\ \ i}$,
respectively. These equivalences preserve the duality between
${\mathbb S}^{A}$ and ${\mathbb S}_{A}$, and between ${\mathbb
S}^{A'}$ and ${\mathbb S}_{A'}$; and at the same time are
consistent with the complex conjugation relations between
${\mathbb S}^{A}$ and ${\mathbb S}^{A'}$, and between ${\mathbb
S}_{A}$ and ${\mathbb S}_{A'}$. Hence if $\alpha^{{\bf A}i}
\in{\mathbb S}^{{\bf A}i}$ then under complex conjugation we have
$\alpha^{{\bf A}i}\to {\bar\alpha}^{{\bf A}'}_{\ \ i}$, and if
$\beta_{{\bf A}i}\in{\mathbb S}_{{\bf A}i}$ then $\beta_{{\bf
A}i}\to{\bar\beta}_{{\bf A}'}^{\ \ i}$.

In the case of a quantum space-time vector $r^{AA'}$ we have a
corresponding induced structure indicated by the identification
\begin{eqnarray}
r^{AA'}=r^{{\bf AA}'i}_{\ \ \ \ \ j}.
\end{eqnarray}
When the quantum space-time vector is real, the weak Hermitian
structure on $r^{AA'}$ is manifested in the form of a standard
\emph{weak} Hermitian structure on the spinor index pair, together
with a \emph{strong} Hermitian structure on the internal index
pair. (In the case of a strong Hermition structure it is assumed
that there is a canonical isomorphism between the complex
conjugate of the given complex vector space and the dual of that
vector space.) In other words, if we define
\begin{eqnarray}
\overline{r^{{\bf AA}'i}_{\ \ \ \ \ \ j}} = {\bar r}^{{\bf A}'{\bf
A}\ \ j}_{\ \ \ \ \ i},
\end{eqnarray}
then the Hermitian condition on the quantum space-time vector
$r^{AA'}$ is given by
\begin{eqnarray}
{\bar r}^{{\bf A}'{\bf A}\ \ i}_{\ \ \ \ \ j} = r^{{\bf AA}'i}_{\
\ \ \ \ j} .
\end{eqnarray}

One striking consequence of these relations is that we can
interpret each point in quantum space-time as being a
\emph{space-time valued operator}. Ordinary classical space-time
then `sits' inside the quantum space-time in a canonical
manner---namely, as the locus of those points of quantum
space-time that factorise into the product of a space-time point
$x^{{\bf AA}'}$ and the identity operator on the internal space:
\begin{eqnarray}
x^{{\bf AA}'i}_{\ \ \ \ \ j} = x^{{\bf AA}'} \delta^{i}_{\ j}.
\label{eq:11.4}
\end{eqnarray}
Thus, in those situations for which special relativity amounts to
a satisfactory theory, we can regard the relevant events as taking
place on or in the immediate neighbourhood of this embedding of
the Minkowski space ${\mathfrak M}^4$ in ${\mathbb R}^{4n^2}$.

This picture can be presented in somewhat more geometric terms as
follows. The hypertwistor space ${\mathbb P}^{2r-1}$ in the case
$r=2n$ admits a Segr\'e embedding of the form ${\mathbb P}^3
\times {\mathbb P}^{n-1}\subset{\mathbb P}^{4n-1}$. Many such
embeddings are possible, though they are all equivalent to one
another under the action of the overall symmetry group $U(2n,2n)$.
If the symmetry is broken and one such embedding is selected out,
then following the conventions discussed earlier we can introduce
homogeneous coordinates and write
\begin{eqnarray}
Z^{\alpha} = Z^{\mba i} .
\end{eqnarray}
Here the bold Greek letter ${\mba}$ denotes an ordinary twistor
index $({\mba}=0.1,2,3)$ and $i$ denotes an internal index
$(i=1,2,\ldots,n)$. The Segr\'e embedding consists of those points
in ${\mathbb P}^{4n-1}$ for which we have a product decomposition
of the associated hypertwistor given by $Z^{\mba i}=
Z^{\mba}\psi^i$.

The idea of symmetry breaking that we are putting forward here is
closely related to the notion of disentanglement in standard
quantum mechanics (cf. Brody \& Hughston 2001). That is to say, at
the unified level the degrees of freedom associated with
space-time symmetry are quantum mechanically entangled with the
internal degrees of freedom associated with microscopic physics.
The phenomena responsible for the breakdown of symmetry are thus
analogous to the mechanisms of decoherence through which quantum
entanglements are gradually diminished.

The compactified complexified quantum space-time ${\mathbb
C}{\mathcal H}_{\sharp}^N = {\mathcal V}^{4n^2}$ can be regarded
as the aggregate of projective $(2n-1)$-planes in ${\mathbb
P}^{4n-1}$. Now generically a ${\mathbb P}^{2n-1}$ in ${\mathbb
P}^{4n-1}$ will not intersect the Segr\'e variety ${\mathcal
G}^{n+2} = {\mathbb P}^{3}\times{\mathbb P}^{n-1}$. Such a generic
$(2n-1)$-plane corresponds to a generic point in ${\mathcal
V}^{4n^2}$. The $(2n-1)$-planes that correspond to the points of
compactified complexified Minkowski space ${\mathbb C}{\mathcal
H}_{\sharp}^{4}$ can be constructed as follows. For each line $L$
in ${\mathbb P}^{3}$ we consider the subvariety ${\mathcal
G}_L^n\subset{\mathcal G}^{n+2}$ where ${\mathcal G}_L^n ={\mathbb
P}_L^1\times{\mathbb P}^{n-1}$. For any algebraic variety
$V^j\subset{\mathbb P}^l$ $(j\leq l-1)$ we define the \emph{span}
of $V^j$ to be the projective plane spanned by the points of
$V^j$. More precisely, we say a point $X$ in the ambient space
${\mathbb P}^l$ lies in the span of the variety $V^j$ if and only
if there exist $m$ points in $V^j$ for some $m\geq2$ with the
property that $X$ lies in the $(m-1)$-plane spanned by those $m$
points. The dimension $k$ of the span of $V^j$ satisfies $j\leq
k\leq l$; however, the value of $k$ depends on the geometry of
$V^j$.

Now the linear span of the points in ${\mathcal G}_L^n$, for any
given $L$, is a $(2n-1)$-plane. This is the ${\mathbb P}_L^{2n-1}$
in ${\mathbb P}^{4n-1}$ that represents the point in ${\mathbb C}
{\mathcal H}_{\sharp}^N$ corresponding to the line $L$ in
${\mathbb P}^3$. The aggregate of such special $(2n-1)$-planes,
defined by their intersection properties with the Segr\'e variety
${\mathcal G}^{n+2}$, constitutes a submanifold of ${\mathcal
V}^{4n^2}$, and this submanifold is compactified complexified
Minkowski space. Thus we see that once the symmetry of quantum
space-time ${\mathcal H}^{4n^2}$ has been broken in the particular
way we have discussed, then ordinary Minkowski space ${\mathfrak
M}^4$ can be identified as a submanifold.

\section{Causality and quantum states}
\label{sec:12}

The embedding of Minkowski space in the quantum space-time
${\mathcal H}^{4n^2}$ given by (\ref{eq:11.4}) implies a
corresponding embedding of the Poincar\'e group in the
hyper-Poincar\'e group. Indeed, if in ${\mathfrak M}^4$ the
standard Poincar\'e group consists of transformations of the form
\begin{eqnarray}
x^{{\bf AA}'}\longrightarrow\, l^{\bf A}_{\bf B}\, \bar{l}^{{\bf
A}'}_{{\bf B}'}\, x^{{\bf BB}'} + b^{{\bf AA}'}, \label{eq:12.1}
\end{eqnarray}
then the hyper-Poincar\'e transformations in ${\mathcal H}^{4n^2}$
are of the form
\begin{eqnarray}
x^{{\bf AA}'i}_{\ \ \ \ \ j}\longrightarrow\, l^{\bf A}_{\bf B}\,
\bar{l}^{{\bf A}'}_{{\bf B}'}\, x^{{\bf BB}'i}_{\ \ \ \ \ j} +
b^{{\bf AA}'}\delta^i_{\,j}. \label{eq:12.2}
\end{eqnarray}
On the other hand, with the identification $A={\bf A}i$, the
general hyper-Poincar\'e transformation in the broken symmetry
phase can be expressed in the form
\begin{eqnarray}
x^{{\bf AA}'i}_{\ \ \ \ \ j}\longrightarrow\, L^{{\bf A}i}_{{\bf
B}k}\, \bar{L}^{{\bf A}'l}_{{\bf B}'j}\, x^{{\bf BB}'k}_{\ \ \ \ \
l} + b^{{\bf AA}'i}_{\ \ \ \ \ j}. \label{eq:12.3}
\end{eqnarray}
Thus the embedding of the Poincar\'e group as a subgroup of the
hyper-Poincar\'e group is given explicitly by
\begin{eqnarray}
L^{{\bf A}i}_{{\bf B}j} = l^{\bf A}_{\bf B}\delta^i_{\ j} \quad
{\rm and} \quad b^{{\bf AA}'i}_{\ \ \ \ \ j} = b^{{\bf AA}'}
\delta^i_{\,j}. \label{eq:12.4}
\end{eqnarray}

Bearing these relations in mind, we now consider the problem of
constructing a certain class of maps from the general
even-dimensional quantum space-time ${\mathcal H}^{4n^2}$ to
Minkowski space ${\mathfrak M}^4$. It will be shown that under
rather general and reasonable physical assumptions such maps
necessarily take the form
\begin{eqnarray}
x^{{\bf AA}'i}_{\ \ \ \ \ j}\longrightarrow\, x^{{\bf AA}'} =
\rho^j_i\,x^{{\bf AA}'i}_{\ \ \ \ \ j}, \label{eq:12.5}
\end{eqnarray}
where $\rho^j_i$ is a \emph{density matrix}, that is to say, a
positive semi-definite Hermitian matrix with unit trace. Thus the
maps under consideration can be regarded as quantum expectations.

\begin{theorem}
Let $\rho:\ {\mathcal H}^{4n^2}\to {\mathfrak M}^4$ satisfy the
following conditions: {\rm (i)} $\rho$ is linear and maps the
origin of ${\mathcal H}^{4n^2}$ to the origin of ${\mathfrak
M}^{4}$; {\rm (ii)} $\rho$ is Poincar\'e invariant; and {\rm
(iii)} $\rho$ is causal. Then $\rho$ is given by a density matrix
on the internal space. \label{theo:2}
\end{theorem}

\begin{proof}
The general linear map from ${\mathcal H}^{4n^2}$ to ${\mathfrak
M}^{4}$ preserving the origin is given by
\begin{eqnarray}
x^{{\bf AA}'i}_{\ \ \ \ \ j}\longrightarrow\, x^{{\bf AA}'} =
\rho^{{\bf AA}'j}_{{\bf BB}'i}\, x^{{\bf BB}'i}_{\ \ \ \ \ j},
\label{eq:12.6}
\end{eqnarray}
where $\rho^{{\bf AA}'j}_{{\bf BB}'i}$ is weakly Hermitian. Now
suppose that we subject ${\mathcal H}^{4n^2}$ to a Poincar\'e
transformation of the form {\rm (}\ref{eq:12.2}{\rm )}, and
require the corresponding transformation of ${\mathcal H}^4$ to be
of the form {\rm (}\ref{eq:12.1}{\rm )}. If $\rho$ satisfies these
conditions then we shall say that the map $\rho$ is Poincar\'e
invariant. It should be evident from this definition that
Poincar\'e invariance holds if and only if
\begin{eqnarray}
\rho^{{\bf AA}'j}_{{\bf BB}'i}\left( l^{\bf B}_{\bf C}\,
\bar{l}^{{\bf B}'}_{{\bf C}'}\,x^{{\bf CC}'i}_{\ \ \ \ \ j} +
b^{{\bf BB}'}\delta^i_{\,j} \right) =  l^{\bf A}_{\bf B}\,
\bar{l}^{{\bf A}'}_{{\bf B}'}\left( \rho^{{\bf BB}'j}_{{\bf CC}'i}
\,x^{{\bf CC}'i}_{\ \ \ \ \ j}\right) + b^{{\bf AA}'}
\label{eq:12.7}
\end{eqnarray}
for all $l^{\bf A}_{\bf B}\in SL(2n,{\mathbb C})$, all $b^{{\bf
AA}'}\in{\mathbb V}^{{\bf AA}'}$, and all $x^{{\bf AA}'i}_{\ \ \ \
\ j}\in{\mathcal H}^{4n^2}$. Thus we have
\begin{eqnarray}
\rho^{{\bf AA}'j}_{{\bf BB}'i}\, l^{\bf B}_{\bf C}\, \bar{l}^{{\bf
B}'}_{{\bf C}'} = l^{\bf A}_{\bf B}\, \bar{l}^{{\bf A}'}_{{\bf
B}'}\, \rho^{{\bf BB}'j}_{{\bf CC}'i}\label{eq:12.8}
\end{eqnarray}
for all $l^{\bf A}_{\bf B}$, and
\begin{eqnarray}
\rho^{{\bf AA}'j}_{{\bf BB}'i}\,\delta^{i}_{\,j}\, b^{{\bf BB}'} =
b^{{\bf AA}'} \label{eq:12.9}
\end{eqnarray}
for all $b^{{\bf AA}'}$. Now {\rm (}\ref{eq:12.8}{\rm )} implies
that $\rho$ is of the form
\begin{eqnarray}
\rho^{{\bf AA}'j}_{{\bf BB}'i} = \delta^{\bf A}_{\,\bf B}\,
\delta^{{\bf A}'}_{\,{\bf B}'}\,\rho^{j}_{i} \label{eq:12.10}
\end{eqnarray}
for some $\rho^j_i$. Then {\rm (}\ref{eq:12.9}{\rm )} implies that
$\rho$ must satisfy the trace condition $\rho^i_i=1$. Finally we
require that if $x^{{\bf AA}'i}_{\ \ \ \ \ j}$ and $y^{{\bf
AA}'i}_{\ \ \ \ \ j}$ are quantum space-time points such that the
interval $r^{{\bf AA}'i}_{\ \ \ \ \ j}=x^{{\bf AA}'i}_{\ \ \ \ \
j}-y^{{\bf AA}'i}_{\ \ \ \ \ j}$ is future-pointing then $r^{{\bf
AA}'}=x^{{\bf AA}'}-y^{{\bf AA}'}$ must also be future pointing,
where $r^{{\bf AA}'}=\rho^j_i\,r^{{\bf AA}'i}_{\ \ \ \ \ j}$. This
is the requirement that $\rho$ should be a \emph{causal} map.
However, this condition immediately implies that $\rho$ must be
positive semi-definite. The argument is as follows. If $r^{{\bf
AA}'i}_{\ \ \ \ \ j}$ is future-pointing then it is necessarily of
the form
\begin{eqnarray}
r^{{\bf AA}'i}_{\ \ \ \ \ j} = \alpha^{{\bf A}i}
\bar{\alpha}^{{\bf A}'}_j + \beta^{{\bf A}i} \bar{\beta}^{{\bf
A}'}_j + \cdots .\label{eq:12.11}
\end{eqnarray}
Consider therefore the case for which $r^{{\bf AA}'i}_{\ \ \ \ \
j}$ is strongly null. Then we require that $\alpha^{{\bf A}i}
\bar{\alpha}^{{\bf A}'}_j\rho^j_i$ should be future-pointing (or
vanish) for any choice of $\alpha^{{\bf A}i}$. Thus in particular
we require that $\alpha^{{\bf A}i} \bar{\alpha}^{{\bf
A}'}_j\rho^j_i$ should be future-pointing if $\alpha^{{\bf A}i}$
is of the form $\alpha^{{\bf A}i}=\alpha^{\bf A} \psi^i$ for any
choice of $\alpha^{\bf A}$ and $\psi^i$. This means that $\rho^j_i
\psi^i\bar{\psi}_j\geq0$ for all $\psi^i$, which shows that
$\rho^j_i$ must be positive semi-definite. Since we have already
shown that the trace of $\rho^j_i$ must be unity, it follows that
$\rho^j_i$ is a density matrix.
\end{proof}

This theorem shows how the causal structure of quantum space-time
is linked in a surprising way with the probabilistic structure of
quantum mechanics. The concept of a quantum state emerges when we
ask for consistent ways of `averaging' over the geometry of
quantum space-time in order to obtain a reduced description of
phenomena in terms of the geometry of Minkowski space.

It is interesting to note that Theorem~\ref{theo:2} has a formal
resemblance to Gleason's theorem in quantum mechanics, which
states that a map from an observable to a real number (expectation
value) must be given by a density matrix, if appropriate
probabilistic conditions are imposed. In the present framework we
see that a probabilistic interpretation of the map from a general
quantum space-time to Minkowski space emerges as a consequence of
elementary causality requirements. We can thus view the space-time
events in ${\mathcal H}^{4n^2}$ themselves as representing quantum
observables, the expectations of which correspond to points of
${\mathfrak M}^{4}$.


\begin{acknowledgements}
DCB gratefully acknowledges financial support from The Royal
Society. The authors are grateful to E.~J.~Brody for numerous
ideas and suggestions in connection with the material presented
here. We are also grateful to J.~Butterfield, D.~Finkelstein,
G.W.~Gibbons, R.~Penrose, and D.C.~Robinson for helpful comments.
\end{acknowledgements}

\end{document}